\documentclass[amssymb,useAMS,prd,aps,preprintnumbers,amsmath,nofootinbib,superscriptaddress,twocolumn]{revtex4-1}
\usepackage{pslatex}
\usepackage{pdfpages}
\usepackage{graphicx}
\usepackage{psfrag}
\usepackage{pdftricks}
\usepackage{multirow}
\usepackage{bigstrut}
\usepackage{hyperref}
\newcommand{\beq}{\begin{equation}}
\newcommand{\eeq}{\end{equation}}
\newcommand{\bea}{\begin{eqnarray}}
\newcommand{\ena}{\end{eqnarray}}

\let\Ga=\Gamma

\let\si=\sigma

\begin{document}

\preprint{ippp/12/20, dcpt/12/40}
\title{Regenerating WIMPs in the light of direct and indirect detection.}

\author{A. J. Williams}
\affiliation{Royal Holloway, University of London, Egham, TW20 0EX, UK}
\affiliation{Particle Physics Department, Rutherford Appleton Laboratory, Chilton,
Didcot, Oxon OX11 0QX, UK}

\author{C. B\oe hm}
\affiliation{Inst. for Particle Physics Phenomenology, Durham University, South Road, DH1 3LE, United Kingdom}
\affiliation{LAPTH, U. de Savoie, CNRS, BP 110, 74941 Annecy-Le-Vieux, France}

\author{S. M. West}
\affiliation{Royal Holloway, University of London, Egham, TW20 0EX, UK}
\affiliation{Particle Physics Department, Rutherford Appleton Laboratory, Chilton,
Didcot, Oxon OX11 0QX, UK}

\author{D. Albornoz V\'asquez}
\affiliation{Institut d'Astrophysique de Paris, UMR 7095 CNRS, Universit´e Pierre et
Marie Curie, 98 bis Boulevard Arago, Paris 75014, France}

\date{17th April 2012}

\begin{abstract}
There are several ways to explain the dark matter relic density other than by the ordinary freeze-out scenario.
For example, the freeze-in mechanism may constitute an alternative for generating the correct relic density for dark matter candidates whose predicted freeze-out abundance is too low due to a large total annihilation cross section. Here we show that although such a mechanism could explain why a dark matter candidate has the correct relic density, some candidates may still be ruled out because 
they would lead to a large gamma ray flux in dwarf spheroidal galaxies or a large elastic scattering rate in direct detection experiments. To investigate this scenario we examine neutralino dark matter in the MSSM. 
However our conclusions can be generalised to other types of annihilating DM candidates with a low relic density in the freeze-out scenario but which have their relic densities generated by some other mechanism.
\end{abstract}
\maketitle

\section{Introduction}
For several decades, the vast majority of dark matter (DM) models have assumed that DM  exists in the form of annihilating particles whose relic density is determined by the freeze-out mechanism \cite{freeze-out1, freeze-out2, freeze-out3}.
This mechanism intimately links the total DM annihilation (or co-annihilation) cross section to the DM relic abundance and leads to the prediction that the total annihilation cross section must be of order the weak scale. 

In some models, nevertheless, the total annihilation cross section at chemical decoupling can be so large (notably if it involves annihilations through resonant channels) that the predicted abundance for the DM candidate is well below the observed value \cite{wmap}, ruling it out as the dominant contribution to the DM relic abundance.  However, new mechanisms, such as the decay of a metastable particle or the freeze-in scenario \cite{Hall:2009bx}, have been proposed as viable alternatives to restore the relic density of such candidates to the required value.

The freeze-in mechanism can be relevant for models where a feeble coupling exists between the thermal bath in the early Universe, consisting of visible sector particles (e.g. standard model or MSSM particles etc), and a thermally decoupled particle species, hereafter referred to as Feebly Interacting Massive Particles (FIMPs). Such a coupling is too small to maintain the FIMPs in thermal equilibrium but it can nevertheless lead to their production via the freeze-in mechanism \cite{Hall:2009bx}. In essence, through some interaction or decay of particles from the visible sector, energy density leaks out of the thermal bath in the form of FIMPs as once they are produced it is unlikely, due to a small number density and the suppressed interactions with the visible sector, that they can reproduce visible sector states, and so their number density accumulates until the process that produces them drops out of equilibrium \cite{Hall:2009bx}.

The freeze-in mechanism can lead to a number of possibilities for DM. For example, if the FIMP is stable (via some stabilising symmetry) it can itself play the role of DM. Its abundance is set directly by the freeze-in mechanism and depends on the size of the interaction or decay rate that leads to the production of FIMP states. 

A second possibility, which we pursue as a working scenario in this paper, is that these FIMPs are unstable and decay into other particles that will constitute the DM. If the FIMP transforms under the same stabilising symmetry that keeps the DM particle stable and has a larger mass, then decays of the FIMP states will generically produce an abundance of DM particles. Due to their small coupling the FIMPs will have a  sufficiently long lifetime such that they will decay after the DM freezes-out and regenerates the DM abundance. In scenarios where the freeze-out abundance of DM is too low, the decay of the frozen-in FIMPs can reproduce the correct abundance\footnote{An appropriate choice for the size of the feeble coupling is needed to get the correct abundance. We treat this coupling as a free parameter and assume that it can be chosen correctly such that the correct final DM abundance is generated.}.

The question that we wish to address in this paper is whether such a mechanism (or related mechanisms) can both explain the observed DM abundance in models where the DM annihilation (or co-annihilation \cite{Griest:1990kh,Binetruy:1983jf}) rate at freeze-out is too large, and be simultaneously compatible with the latest results from direct and indirect detection experiments. In particular, for direct detection constraints, we apply limits on the DM-nucleon spin-independent elastic cross section as derived from the XENON100 experiment \cite{Aprile:2011hi}. The indirect detection limits that we apply come from the latest observations of the dwarf spheroidal galaxies (dSphs) by FERMI-LAT \cite{Abdo:2010ex}, which place an upper limit on the gamma flux emerging from DM annihilations.

We consider as our DM  candidate the neutralino of the MSSM and assume that one can add an extra term or terms to the MSSM Lagrangian in order to implement the freeze-in scenario\footnote{See \cite{Hall:2009bx, Cheung:2010gk, Cheung:2010gj} for examples.}. The use of Supersymmetry enables us to explore very different types of configurations in terms of resonances and co-annihilations. However, similar conclusions will also hold for other types of DM candidates where the freeze-out relic density is too low due to a large annihilation cross section.
 
In Section \ref{paramspace}, we describe our method for investigating the parameter space of the MSSM with the aim of determining the regions that lead to an annihilation cross section greater than the common value of $\sigma v \approx 3\times 10^{-26}{\rm cm}^3{\rm s}^{-1}$, which is needed in the standard freeze-out scenario. This is indeed where freeze-in (and similar mechanisms) can be important as regions of parameter space that were previously unable to explain why the DM candidate would make up all of the WMAP observed value are now potentially viable. In Section \ref{DM_regeneration}, we investigate the phenomenology of these ``under abundant" configurations in the light of DM experiments and conclude in Section \ref{conclusions}.
\begin{center}
\begin{table}
\renewcommand{\arraystretch}{1.1}
\begin{tabular*}{\columnwidth}{@{\extracolsep{\fill}}|c|@{\extracolsep{0pt}}c|}
\cline{1-2}
Scan A &  Scan B \\ \cline{1-2}
$2\;{\rm GeV} <M_1<120 $ GeV & $90\;{\rm GeV} <M_1<2000 $ GeV \\ \cline{1-2}
  \multicolumn{2}{|c|}{$90\;{\rm GeV} <M_2<2000 $ GeV } \\ \cline{1-2}
  \multicolumn{2}{|c|}{$200\;{\rm GeV} <M_3<6000 $ GeV} \\ \cline{1-2}
  \multicolumn{2}{|c|}{$2\;{\rm GeV} <\mu<2000 $ GeV} \\ \cline{1-2}
 \multicolumn{2}{|c|}{$0.1<\tan \beta<75$} \\ \cline{1-2}
 \multicolumn{2}{|c|}{$-4000\;{\rm GeV} <A_t<4000 $ GeV} \\ \cline{1-2}
 \;\; $100\;{\rm GeV} <m_{A^0}<1500 $ GeV\;\; &\;\; $100\;{\rm GeV} <m_{A^0}<4000$ GeV\;\; \\ \cline{1-2}
 \multicolumn{2}{|c|}{$100\;{\rm GeV} <m_{\tilde{l}_L}<4000 $ GeV}\bigstrut[b] \\ \cline{1-2}
 \multicolumn{2}{|c|}{$100\;{\rm GeV} <m_{\tilde{l}_R}<4000 $ GeV} \bigstrut[b]\\ \cline{1-2}
 \multicolumn{2}{|c|}{$100\;{\rm GeV} <m_{\tilde{q}_{1,2}}<4000 $ GeV} \\ \cline{1-2}
 \multicolumn{2}{|c|}{$100\;{\rm GeV} <m_{\tilde{q}_{3}}<4000 $ GeV} \\ \cline{1-2}
\end{tabular*}
\caption{Allowed ranges of the parameters.\label{tab:parameters}}
\end{table}
\end{center}
\section{Parameter space}\label{paramspace}
We consider the MSSM and allow for 11 parameters to vary, namely the gaugino masses, $M_1$, $M_2$, $M_3$, the Higgs-Higgsino mass parameter, $\mu$, the ratio of the Higgs vacuum expectation values, $\tan \beta$, the stop trilinear coupling, $A_t$, (all other trilinear couplings are set to zero), the mass of the CP-odd Higgs, $m_{A^0}$,  and finally the parameters $m_{\tilde{q}_{1,2}}$, $m_{\tilde{q}_{3}}$ and $m_{\tilde{l}_{L,R}}$, which represent the squark masses for the first two generations, the third generation squark masses, and all generations of  the ``left" and ``right" sleptons respectively. 

The choice of $m_{\tilde{q}_{1}} = m_{\tilde{q}_2} \neq m_{\tilde{q}_3} $ is particularly relevant since stops can be lighter than the first two generations and can be relevant for enhancing neutralino annihilations (cf e.g. \cite{Boehm:1999bj}). Separating the ``left" and ``right" slepton masses also allows for a light slepton (mostly in the case of ``left" sleptons) that can play a significant role in neutralino co-annihilations \cite{Ellis:1998kh,Ellis:1999mm}. The values of all parameters are defined at the electroweak scale. 

We perform two Markov Chain Monte Carlo (MCMC) scans labelled Scan A and Scan B. Scan A is dedicated to low neutralino masses (below 100\;GeV) while Scan B is dedicated to heavier candidates (above 100\;GeV). This choice of two separate scans above and below 100\;GeV is purely arbitrary but it turns out to be a useful division. The reason being that the neutralino candidates found in each scan represent different freeze-out scenarios and are most sensitive to different experimental searches. Scan A features s-channel resonant effects while Scan B shows a greater number of t-channel exchange and co-annihilation processes. In these scans we are looking for points where the freeze-out relic density, $\Omega_{\rm FO}h^2$, is lower than the mean value $\Omega_{\rm WMAP} h^2 \simeq 0.1123$ (obtained by combining WMAP data with BAO and $ H_0$ measurements \cite{wmap7yr}) and which satisfy a number of constraints from particle physics experiments. We take the mean value for $\Omega_{\rm WMAP} h^2$ as an absolute limit with no uncertainty rather than the WMAP maximal upper bound as we are interested in only those points that would require regeneration in order to fit the WMAP data. Since the most interesting region of the parameter space in terms of regeneration is far away from the limit, the uncertainty in the mean value can be neglected without affecting the results.
Our choices for the allowed ranges of the MSSM parameters for the two scans are listed in Table~\ref{tab:parameters} and the constraints used to calculate likelihoods can be found in Table~\ref{tab:constraints}.

\begin{table}
\begin{tabular*}{\columnwidth}{|c|c|c|}
\hline
Constraint & Value & Tolerance \\ \hline
$\Omega_{\rm FO} h^2$ & $< 0.1123$ \cite{wmap7yr} & none \\ \hline
$(g - 2)_{\mu} $ & $25.5 \times 10^{-10}$ & \hspace{1mm} stat: $6.3 \times 10^{-10}$ \\
  & &\hspace{1mm} sys: $4.9 \times 10^{-10}$ \\ \hline
 $\Delta \rho$ & $\leq 0.002$ & $0.0001$ \\ \hline
 BF$(b \rightarrow s \gamma)$ &\hspace{2mm} $3.55 \times 10^{-4}$\; \cite{HFAG2010}\hspace{5mm} &\hspace{5mm} th: $0.24 \times 10^{-4}$ \\ \hline
 BF$(B_s \rightarrow \mu^+ \mu^-)$ & $\leq 4.5 \times 10^{-9}$ \cite{Aaij:2012ac} & $4.5 \times 10^{-11}$ \\ \hline
 R$(B \rightarrow \tau \nu_\tau)$ & $1.36\;$  \cite{HFAG2010} & $0.23$ \\ \hline
 $\Ga(Z \rightarrow \tilde{\chi}^0_1 \tilde{\chi}^0_1)$ & $\leq 1.7\;\text{MeV}$ & $0.3\;\text{MeV}$ \\ \hline
 $\si(e^+ e^- \rightarrow \tilde{\chi}^0_1 \tilde{\chi}^0_{2,3})$ \hspace{1.6mm} & $\leq 0.1\; \text{pb}\;$ \cite{Abbiendi:2003sc}   & $0.001 \text{pb}$\\ \hline
\end{tabular*}
\caption{Constraints used to calculate likelihoods, from Ref.~\cite{Nakamura:2010zzi} unless stated. Here $\Omega_{\rm FO} h^2$ is the relic abundance of neutralino DM from freeze-out, $\Delta \rho$ is the contribution to the electro-weak precision variable $\rho$, R$(B \rightarrow \tau \nu_\tau)$ is the ratio of the MSSM to SM branching fraction of $B^+ \rightarrow \tau^+ \nu_\tau$.}
\label{tab:constraints}
\end{table}

All the physical quantities in this analysis are computed using the micrOMEGAs code \cite{Belanger:2001fz} except for the SUSY particle spectrum and decay rates of the Higgs particles which were calculated using SoftSusy \cite{softsusy} and SUSYHIT \cite{susyhit} respectively.  LEP limits on the sparticle masses are applied automatically by micrOMEGAs (see \cite{Belanger:2001fz} for details). In addition, a lower limit on the squark masses is set at 100\;GeV. The Higgs masses are restricted to the allowed range by the HiggsBounds programme \cite{higgsbound1, higgsbound2}. SoftSusy, SUSYHIT, micrOMEGAs and HiggsBounds were interfaced via the SUSY Les Houches Accord \cite{Allanach:2008qq}. We do not apply limits on the squark and gluino masses coming from the latest CMS and ATLAS data. The effect of these limits could be considered by simulating events for a converged sub set of the Markov chains as described in \cite{Sekmen:2011cz}, however, we consider this beyond the scope of this particular work. Our focus here is to examine the possibility of regenerating the DM density in under-abundant DM scenarios in the light of DM experiments.  

To explore the parameter space we generate a random walk using the Metropolis-Hastings algorithm. For these scans a an initial point in parameter space is randomly chosen. Following this steps in the random walk are taken along randomly selected directions in the parameters space and an initial ``burn-in" phase is used to adjust the magnitude of the proposed step size for each direction to optimise the exploration of the parameter space, this is periodically adjusted during ``burn-in" to ensure that the parameter space is covered as fully as possible. The directions in which steps are taken were generated from the eigenvectors of the covariance matrix found in preliminary scans. The ``burn-in" phase also ensures that the chain has already converged towards a high likelihood before points are recorded. The total likelihood function is formed by the product of partial likelihoods for each observable in Table~\ref{tab:constraints}. As in Ref.~\cite{Vasquez:2010ru} we use a Gaussian distribution for observables with a preferred value
\begin{equation}
F_2(x,\mu ,\sigma) = e^{-\frac{(x - \mu)^2}{2\sigma^2}},
\end{equation}
where $\mu$ is the preferred value of the observable and $\sigma$ is the tolerance. For observables with only an upper or lower limit a distribution of the form,
\begin{equation}
F_3(x,\mu ,\sigma) = \frac{1}{1 + e^{-\frac{(x - \mu)}{\sigma}}},
\end{equation}
is used. Here $\sigma$ is positive for lower bounds and negative for upper bounds. For the relic abundance, the masses of the sparticles and the Higgs masses, the partial likelihood is either one or zero as no uncertainties are included.

\begin{table}
\renewcommand{\arraystretch}{1.2}
\begin{tabular*}{\columnwidth}{|c|c|c|}
\hline
Standard model parameter & \hspace{0.4mm} Mean value \hspace{0.3mm}& Experimental uncertainty \\ \hline
$m_t$ & $172.9$\;GeV & $1.5$ \\ \hline
$m_b(m_b)^{\overline{MS}}$ & $4.19$\;GeV & $^{+0.18}_{-0.06}$ \\ \hline
$\alpha_s(m_Z)^{\overline{MS}} $ & 0.1184 & 0.0007 \\ \hline
$\alpha^{-1}_{EM}(m_Z)^{\overline{MS}} $ & $127.916$ & $0.015$ \\ \hline
\end{tabular*}
\caption{Constraints used to calculate likelihoods for standard model parameters, from Ref.~\cite{Nakamura:2010zzi}.}
\label{tab:nuisance}
\end{table}

Uncertainties in standard model parameters were included in the form of nuisance parameters which are then marginalised as part of the random walk. The mean values and uncertainties of the nuisance parameters are shown in Table~\ref{tab:nuisance}.

\subsection{Scan A: results for scenarios with $m_{\tilde{\chi}_1^0} <$ 100 GeV}
\begin{figure}[t]
\includegraphics[width=9cm]{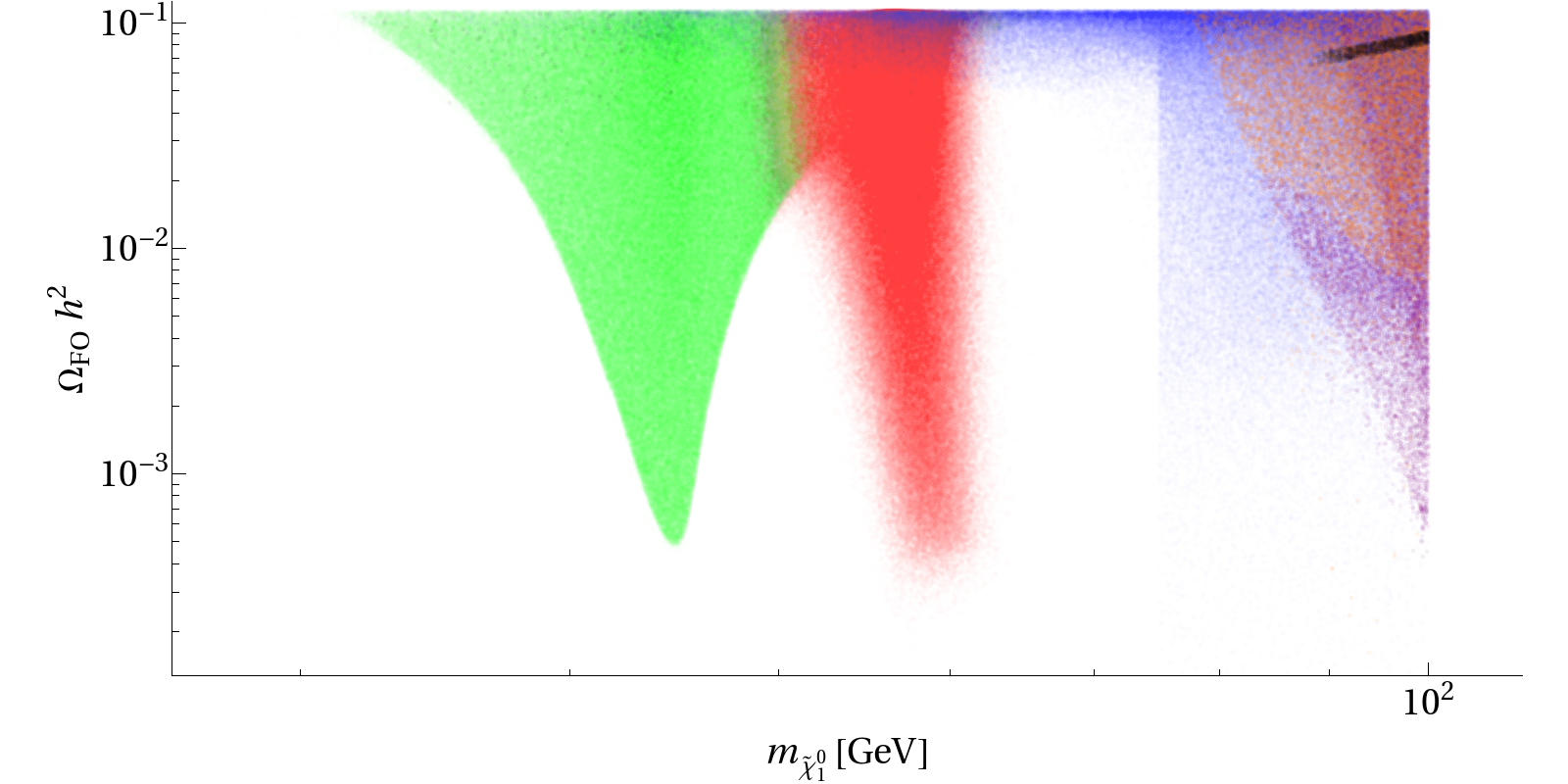}
\caption{Plot of $\Omega_{\rm FO} h^2$ against $m_{\tilde{\chi}^0_1}$. The colour coding represents the process with the largest contribution to the neutralino annihilation rate, which determines the freeze-out relic abundance. Green points correspond to resonant annihilation via Z, red points to resonant annihilation via the light Higgs boson ($h^0$), orange points to resonant annihilation via the pseudo-scalar Higgs ($A^0$), blue points to stau co-annihilation or annihilation via stau exchange, violet points to chargino co-annihilations or chargino exchange, black points to squark co-annihilation (all squark flavours).}
\label{fig:channels_below100GeV}
\end{figure}

Previous supersymmetric parameter scans either looked for scenarios with the correct relic density (e.g. \cite{Griest:1991gu,Griest:1989zh,Drees:1992am,Roszkowski:2001sb,Buchmueller:2011ki,Ellis:2007ka,Barger:2001yy,Ellis:2012aa}) or relaxed the 
constraint on the relic density, allowing for very small $\Omega_{\rm FO} h^2$, and did not assume the presence of regeneration mechanism \cite{Vasquez:2010ru,AlbornozVasquez:2011yq}. In this paper we will both relax the lower bound on the relic density and assume that the freeze-in mechanism can regenerate the relic density to the observed value.

In FIG.~\ref{fig:channels_below100GeV}, we show the relic density versus DM mass for candidates found by the MCMC. In most scenarios more than one process will contribute to the freeze-out relic abundance but in FIG.~\ref{fig:channels_below100GeV} the largest single contribution to the annihilation rate, which in the majority of scenarios dominates the others, is indicated. In all of the following plots the points found by the random walk are plotted as semi-transparent dots, faint regions therefore correspond to a low density of points while regions of strong colour correspond to denser regions. As expected there are two visible resonance regions \cite{Griest:1990kh}, corresponding to Z gauge boson and light CP-even Higgs ($h^0$) s-channel resonances. In addition there are the usual points corresponding to heavier neutralinos that can annihilate via s-channel exchange of the CP-odd Higgs ($A^0$) \cite{Roszkowski:2001sb}, as is well known from traditional freeze-out scenarios. These points appear as a smeared out region due to the large variation in the value of $m_{A^0}$. 

In addition to the s-channel processes the well known t-channel exchange and co-annihilations processes involving charginos, staus and squarks are also found by the MCMC. It is likely that the majority of the points corresponding to squark exchange and co-annihilation will be excluded by the LHC or Tevatron. However, we still include these points as our focus here is to examine the effect of regeneration and the resulting DM detection constraints on the possible regions of the parameter space.

The composition of the neutralino LSP in terms of the weak eigenstates, the Bino, Higgsinos  and Wino differs slightly for the various regions displayed in FIG.~\ref{fig:channels_below100GeV}.

For the Z and $h^0$ resonance regions the neutralino is mostly Bino with a small Higgsino component. As is well known, (see for example \cite{Feng:2000gh,Feng:2011aa}), the size of the Higgsino component will play a central role in determining the cross section for DM annihilations via s-channel Z and $h^0$. This Higgsino component will also lead to the dominant  contributions to the spin-independent elastic scattering cross section in direct detection experiments, where the main process is the t-channel exchange of a Higgs. This connection is important for what follows in the later sections. 

In the cases where t-channel exchange and co-annihilation processes, involving light SUSY squarks and sleptons, dominate the freeze-out dynamics, the neutralino can have a much smaller Higgsino component. This is because, in contrast to the s-channel annihilation processes, the t-channel annihilation and co-annihilation diagrams can occur for pure Bino neutralinos.

\subsection{Scan B: results for scenarios with $m_{\tilde{\chi}_1^0} >$ 100 GeV}

\begin{figure}[t]
	\centering
\includegraphics[width=9cm]{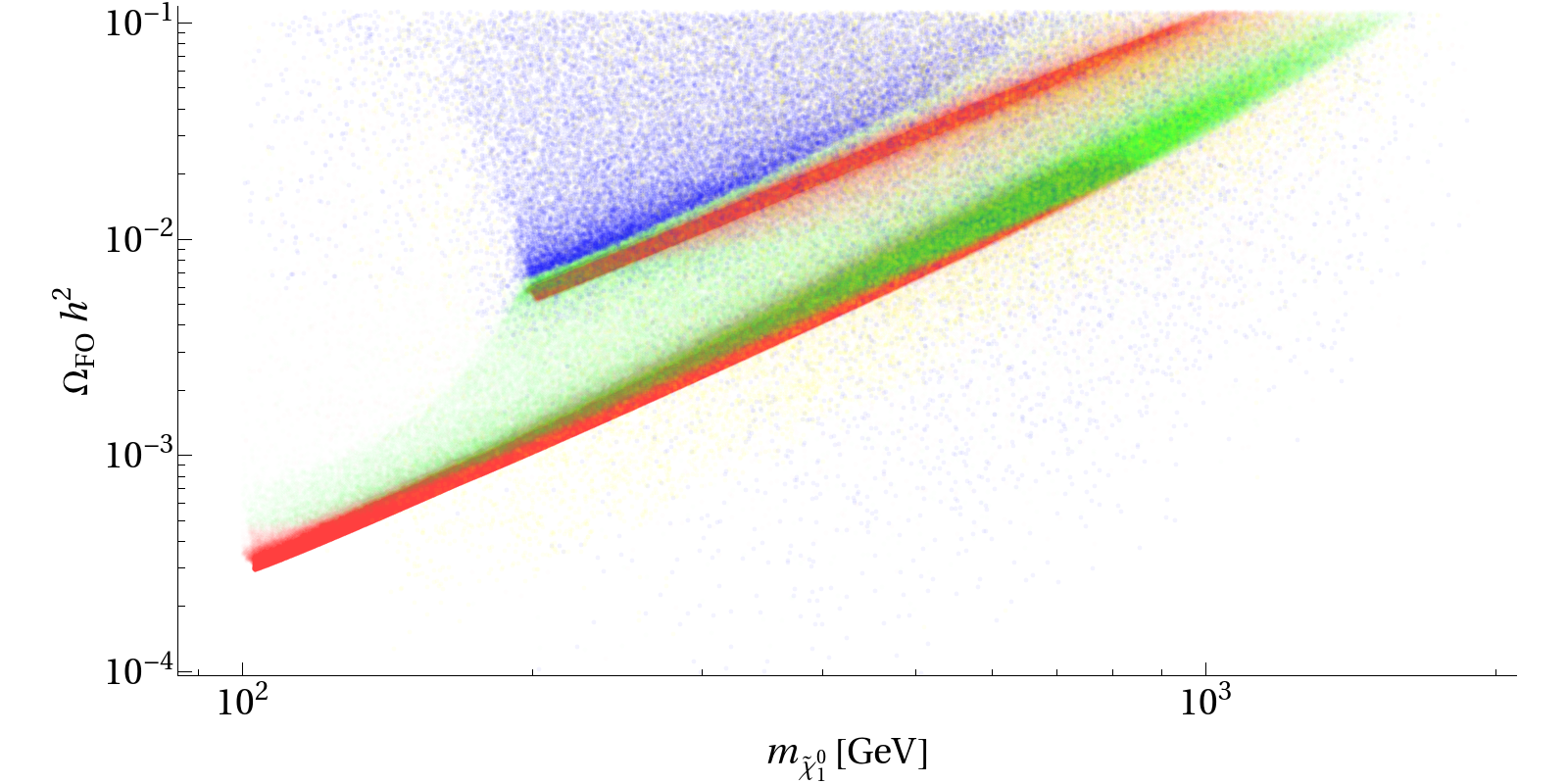}
\caption{Plot of $\Omega_{\rm FO} h^2$ against $m_{\tilde{\chi}^0_1}$. Colour coded for the process with the largest contribution to the total neutralino annihilation rate, which determines the freeze-out relic abundance. Red points correspond to chargino co-annihilation, green points to annihilation via chargino t-channel exchange, blue points to annihilation via s-channel Higgs (roughly speaking the blue points above the green band correspond to annihilation via an s-channel $h^0$ into $t\bar{t}$ and $b\bar{b}$, the few below are s-channel annihilation via $A^0$), yellow points correspond to a either squark co-annihilation or gluino-gluino annihilations (the latter in the case where the gluino is approximately mass degenerate with the neutralino DM and its freeze-out sets the neutralino relic abundance).}
\label{fig:channels_above100GeV}
\end{figure}

In the case of neutralinos heavier than 100\;GeV, one does not expect any resonance structure in the $(m_{\tilde{\chi}_1^0},\Omega_{\rm FO} h^2)$ plane since there are no fixed mass neutral particles (such as the light CP-even Higgs\footnote{Although the $h^0$ mass is not fixed, it is restricted to a narrow range in the MSSM.} or Z boson) that can be produced in an s-channel resonance. Instead resonant annihilation through\;$A^0$ will appear over a range of different neutralino masses. Non-resonant annihilation via the $h^0$ and Z bosons can still produce a large enough cross section to reduce the relic abundance for masses above 200\;GeV.  Chargino or squark t-channel exchange and co-annihilations also lead to an enhanced cross section but this does not appear as a fixed mass resonance. As a result, we find a smooth homogeneous distribution of points in the $(m_{\tilde{\chi}_1^0},\Omega_{\rm FO} h^2)$ plane, as shown in FIG.~\ref{fig:channels_above100GeV}.

The most visible trend in FIG.~\ref{fig:channels_above100GeV} is that the minimum relic abundance found by the MCMC increases quadratically as a function of mass. This dependence of the relic abundance on the mass of the neutralino DM arises due to the fact that the relic abundance scales as the inverse of the thermally averaged cross section, which in turn scales approximately as the inverse of the neutralino mass squared. As a result the minimum relic abundance will increase quadratically with the mass of the neutralino.  Co-annihilation with light stops is expected to add a few more points (below the ``quadratic" limit) when there is a large fine-tuning between the neutralino and the stop mass. However, the stop and neutralino self-annihilation cross sections both decrease with the mass of these particles and an increase in the fine tuning becomes less and less effective in compensating for the lack of efficiency of the co-annihilation process when the neutralino mass increases.  Besides, these points become more difficult to find by the MCMC as they require smaller variance (i.e. more dedicated searches).

The compositions of the higher mass neutralinos is more varied than the lower mass states. For example, in points whose freeze-out annihilation rate is dominated by chargino co-annihilation and t-channel chargino exchange the neutralino DM can be mostly Wino. For points whose freeze-out annihilation is dominated by s-channel Higgs processes, the Higgsino component of these neutralinos can be much larger (even dominating the composition) than that for neutralino DM with masses below $100\;$GeV.

\section{DM regeneration in the light of FERMI-LAT and XENON100 limits}\label{DM_regeneration}

 \begin{figure*}[t]
	\centering
\includegraphics[width=0.49\textwidth]{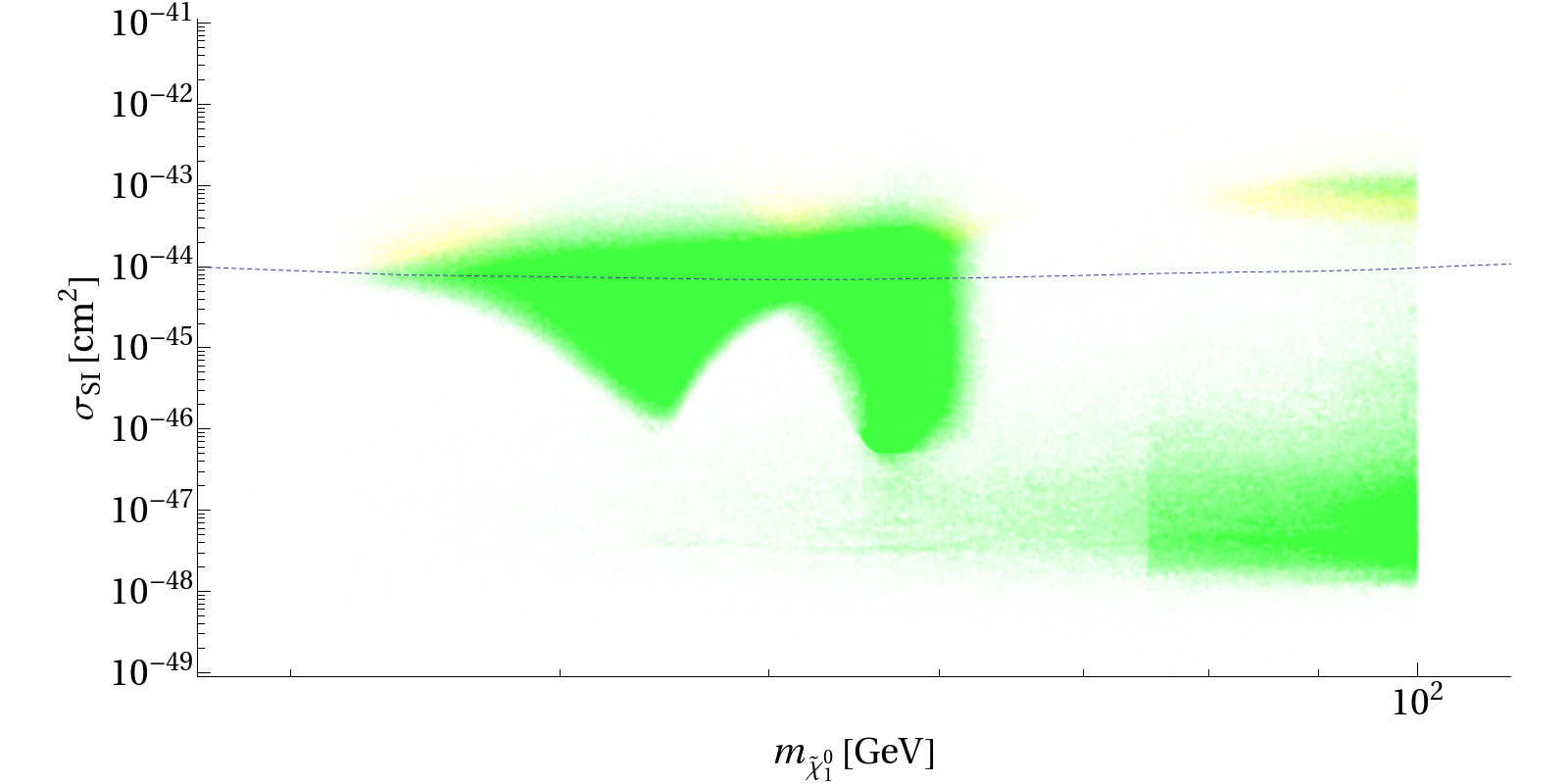}
\includegraphics[width=0.49\textwidth]{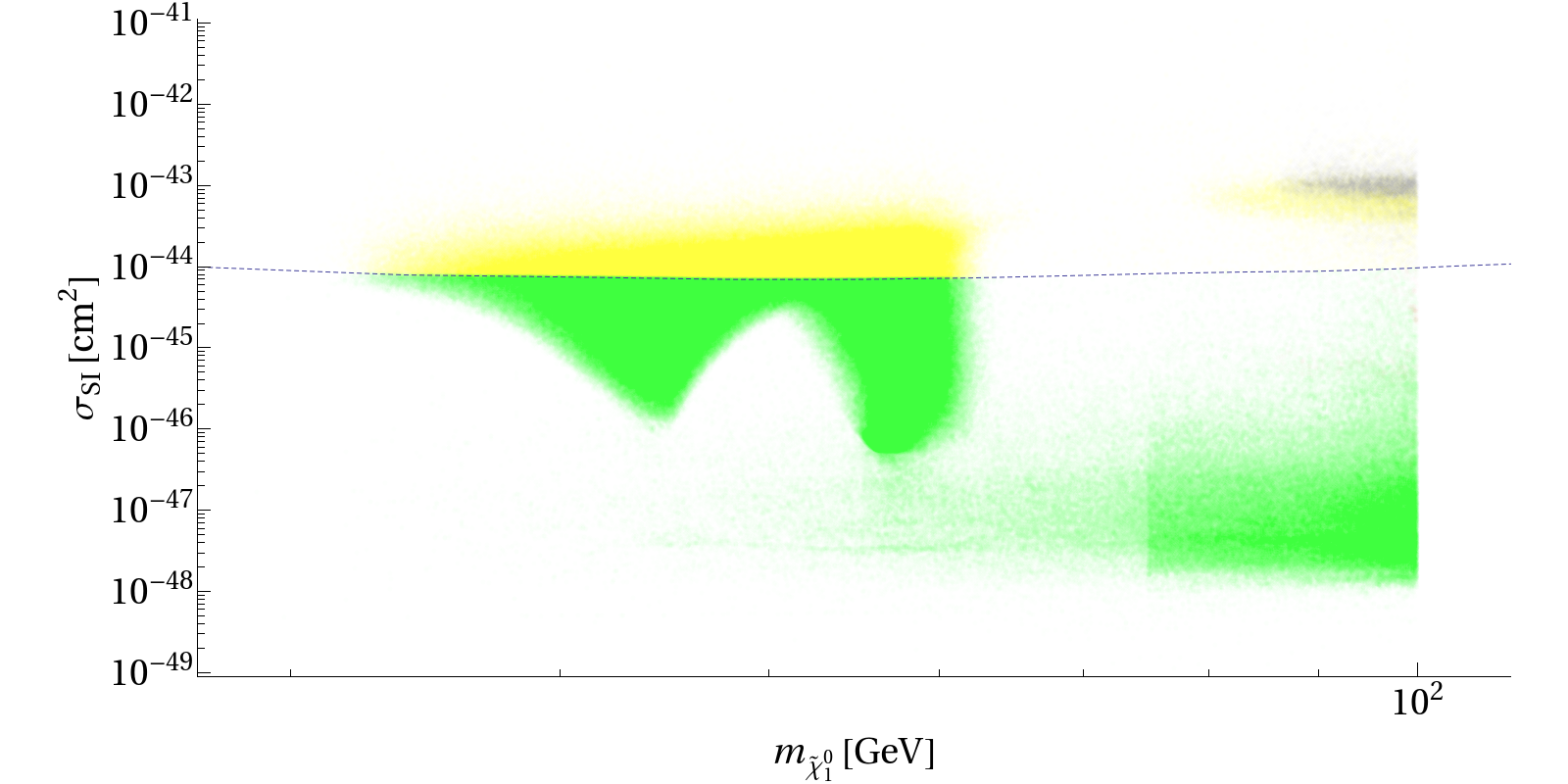}
\caption{Spin-independent cross section versus neutralino mass. The right panel shows the case with the regeneration of the DM relic density  to the correct value, the left panel shows the case without. The limit from XENON100 \cite{Aprile:2011hi} as a limits on the spin-independent cross section of a DM species with the WMAP observed relic density is shown as a blue dashed line. The yellow points are excluded by XENON100, the red points are excluded by indirect detection, grey points are excluded by both and green points survive all constraints applied.}
\label{fig:direct}
\end{figure*}

To examine the impact of a possible regeneration mechanism we apply limits arising in direct and indirect detection experiments to the points found by the MCMC. We do so in two cases. The first where there is no regeneration and the DM density is set by the value determined by freeze-out. The second where regeneration of the DM density has taken place after freeze-out and has been regenerated to the WMAP observed value. The limits for direct and indirect detection are applied as 95\% confidence level exclusions to the points found by the MCMC after the scans have completed rather than including these limits in the likelihood calculations. This allows the two scenarios to be compared directly using the same set of points.

We look at the effect of regeneration in the planes ($\sigma_{\rm SI}, m_{\tilde{\chi}_1^0}$), ($\sigma_{\rm SI}, \Omega_{\rm FO}h^2$), ($\Phi_{\rm PP}, m_{\tilde{\chi}_1^0}$) and ($\Phi_{\rm PP}, \sigma_{\rm SI}$), where $\sigma_{\rm SI}$ is the spin-independent elastic scattering rate, $\Omega_{\rm FO}h^2$ is the relic abundance generated by freeze-out only and $\Phi_{\rm PP}$, which encodes the ``particle physics input" to the total flux of gamma rays from annihilating DM in the dSphs. The quantity $\Phi_{\rm PP}$ is defined as
\begin{equation}
\Phi_{\rm PP} = \frac{\langle \sigma v \rangle}{8 \pi m_{\tilde{\chi}^0_1}^2} \int_{E_0}^{E_{max}} \frac{dN}{dE} dE,
\end{equation}
where $\langle \sigma v \rangle$ is the thermally averaged cross section for DM annihilation, $E_0$ is the minimum threshold energy considered, $E_{max}$ is the maximum photon energy the limit is sensitive to and $\frac{dN}{dE}$ is the gamma ray spectrum averaged over all of the different annihilation channels. Neglecting propagation the expected flux of gamma rays from a given source reads as
\begin{equation}
\Phi_{\gamma} = \Phi_{\rm PP} \times J,
\end{equation}
where $J$ is the DM density integrated along the line of sight and over the solid angle and sensitivity of the observation. 

An upper limit on the flux and a particular choice of $J$ then set an upper limit on $\Phi_{\rm PP}$. In general the upper limit on the flux depends on assumptions about the spectral shape of the gamma ray source. Choosing the hardest power-law model from \citep{Abdo:2010ex} gives an upper bound on the photon flux which can be divided by $J$ to give a conservative upper bound of $\Phi_{\rm PP} < 7.5 \times 10^{-30} \text{cm}^3 \text{s}^{-1} \text{GeV}^{-2}$ from observations of the Draco  dSph by FERMI-LAT \cite{Abdo:2010ex}. However, using a combined analysis of several dSphs places a stronger limit of $\Phi_{\rm PP} < 5.0 \times 10^{-30} \text{cm}^3 \text{s}^{-1} \text{GeV}^{-2} $\cite{GeringerSameth:2011iw}. In this case there is no single limit on the gamma ray flux and corresponding $J$ value, instead the limit on $\Phi_{\rm PP}$ is found by Neyman construction \cite{Feldman:1997qc,neyman} where each dSph is weighted by its $J$ value. We use this combined limit in what follows.  

For each point found by the MCMC the gamma ray spectrum $\frac{dN}{dE}$ is calculated using micrOMEGAs and integrated from $1$\:GeV to $100$\;GeV in order to obtain $\Phi_{\rm PP}$.

In addition to applying constraints from indirect detection, we also apply constraints coming from direct detection experiments. In particular we apply the limits on the spin-independent elastic cross section coming from XENON100 \cite{Aprile:2011hi}. The spin-independent cross section for each point is calculated automatically in micOMEGAs and we refer the reader to \cite{Belanger:2001fz} for details. One important point we do note here is that we use the default values for the scalar form factors of the proton and neutron as set in micOMEGAs \cite{Belanger:2001fz}. In particular we use the default value for the strange quark scalar form factors as given in \cite{Belanger:2001fz} as $f_s^{n,p}=0.2594$. It is well known that this is a source of a large uncertainty in direct detection rates, see for example \cite{Ellis:2008hf,Buchmueller:2011ki,Giedt:2009mr} and can lead to a significant change in the predicted cross sections. Astrophysical uncertainties can also have an impact on the limits applied, see \cite{Green:2010ri,McCabe:2010zh} but again we do not allow for these uncertainties.  

Finally, for $m_{\tilde{\chi}_1^0}<$ 50 GeV, the uncertainties on the exclusion curve, due to the lack of physical knowledge on the energy behaviour of the relative scintillation efficiency, are important. The latter do not appear  in \cite{Aprile:2011hi} because the XENON100 collaboration assumed that the uncertainties on the relative scintillation efficiency can be well modelled by a Gaussian likelihood  centred on the ${\cal{L}}_{\rm{eff}}$ mean value. It was not realised that maximising the global likelihood gives more weight to the mean (but not necessarily the physical) value of ${\cal{L}}_{\rm{eff}}$ and does not allow the real (physical) uncertainties on ${\cal{L}}_{\rm{eff}}$ to be taken properly into account \cite{Davis:2012vy}. Here we continue to use the exclusion curve obtained in \cite{Aprile:2011hi} as a guideline to understand the effect of regeneration but a more detailed study would require the implementation of all these sources of uncertainties in the derivation of the direct detection exclusion limit.

In the following subsections we present a series of double panel figures. The plots corresponding to no regeneration (freeze-out only contributions to the DM relic density) are displayed in the left panels. The same points are plotted in the right panels but now with the DM density regenerated to the WMAP observed value.  Note that these scenarios are strictly identical in the pairs of plots apart from the DM densities used to calculate the limits. It should be noted that in the calculations for the indirect detection rates, micrOMEGAS \cite{Belanger:2001fz} uses by default the value of the DM density determined by WMAP \cite{wmap} not the value predicted by freeze-out, which in the majority of our cases will be below the WMAP value. In order to calculate the gamma ray flux for the under-abundant scenarios, the square of the scaling factor, $\eta$, needs to be applied, where $\eta=\Omega_{\rm FO}/\Omega_{\rm WMAP}$. Similarly, for the under-abundant scenarios, the limits on the elastic scattering cross section from direct detection need to be scaled by $\eta$.

\begin{figure*}[t]
	\centering
\includegraphics[width=0.49\textwidth]{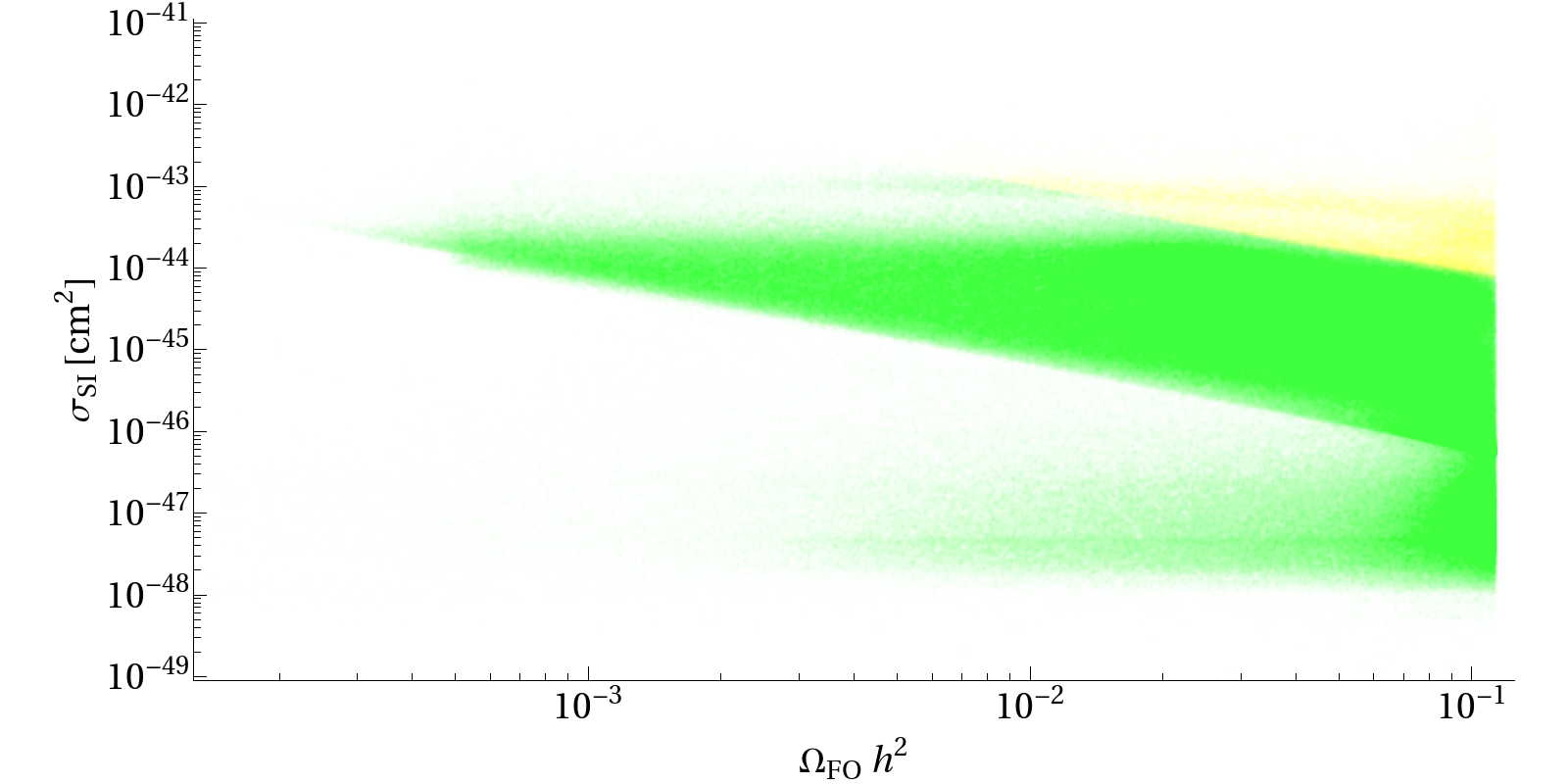}
\includegraphics[width=0.49\textwidth]{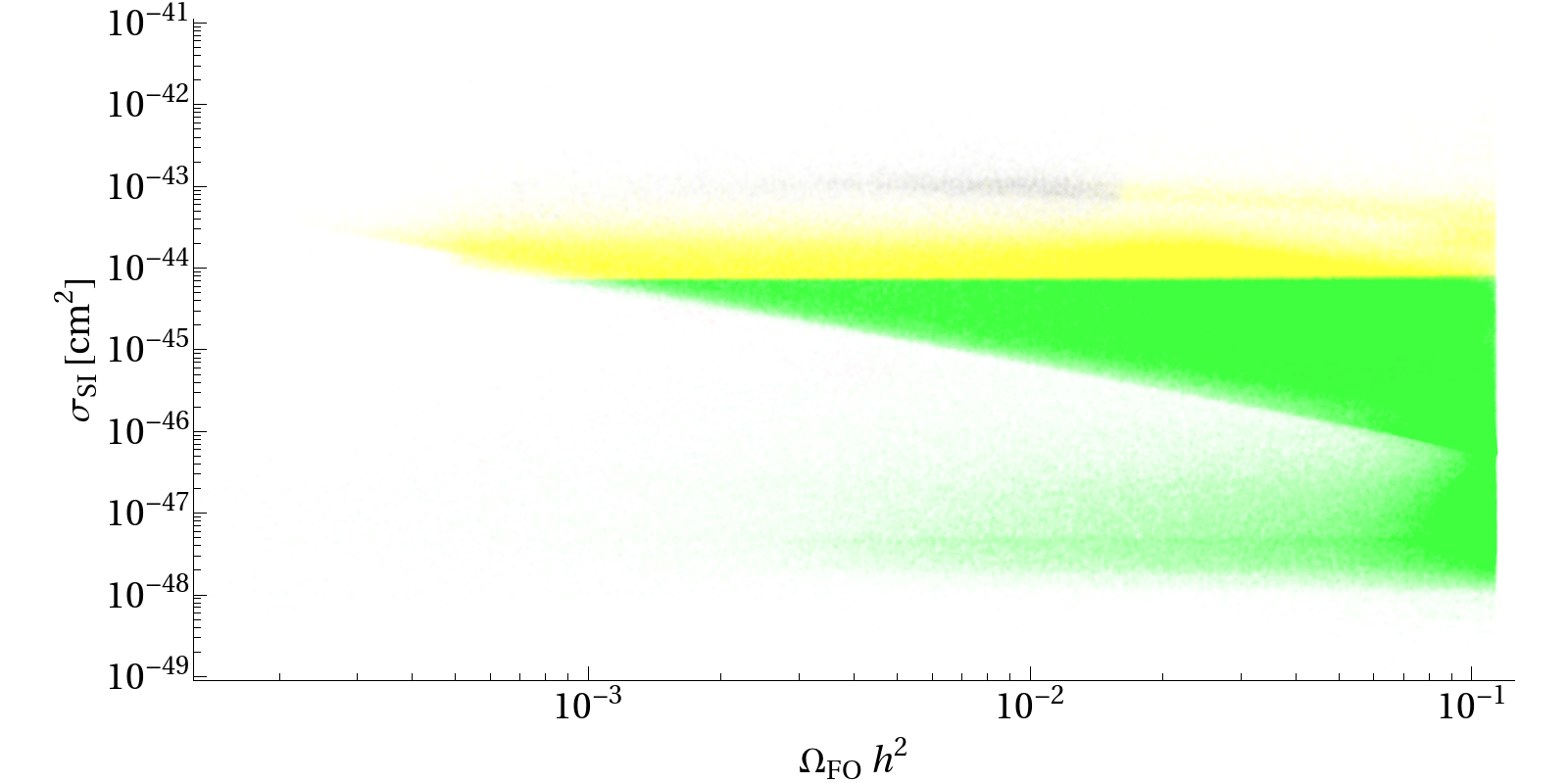}
\caption{Spin-independent cross section versus the neutralino freeze-out relic density for $m_{\tilde{\chi}_1^0} < 100$ GeV. The right panel shows the case with the regeneration of the DM relic density  to the correct value, the left panel shows the case without. Colour coding is the same as in FIG.~\ref{fig:direct}.}
\label{fig:RD}
\end{figure*}

\subsection{Regeneration in scenarios with $m_{\tilde{\chi}_1^0} <$ 100 GeV}

 \begin{figure*}[t]
	\centering
\includegraphics[width=0.49\textwidth]{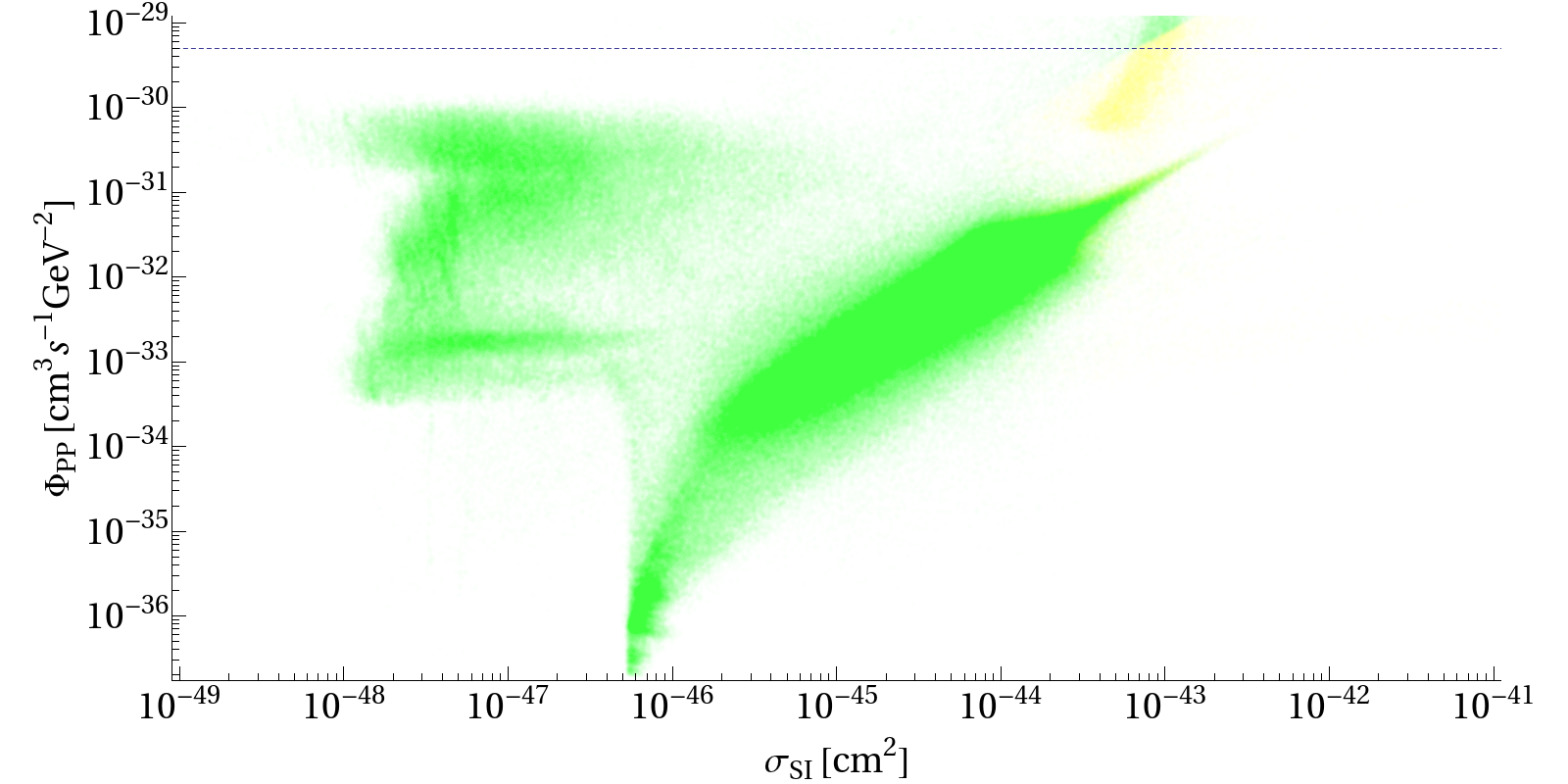}
\includegraphics[width=0.49\textwidth]{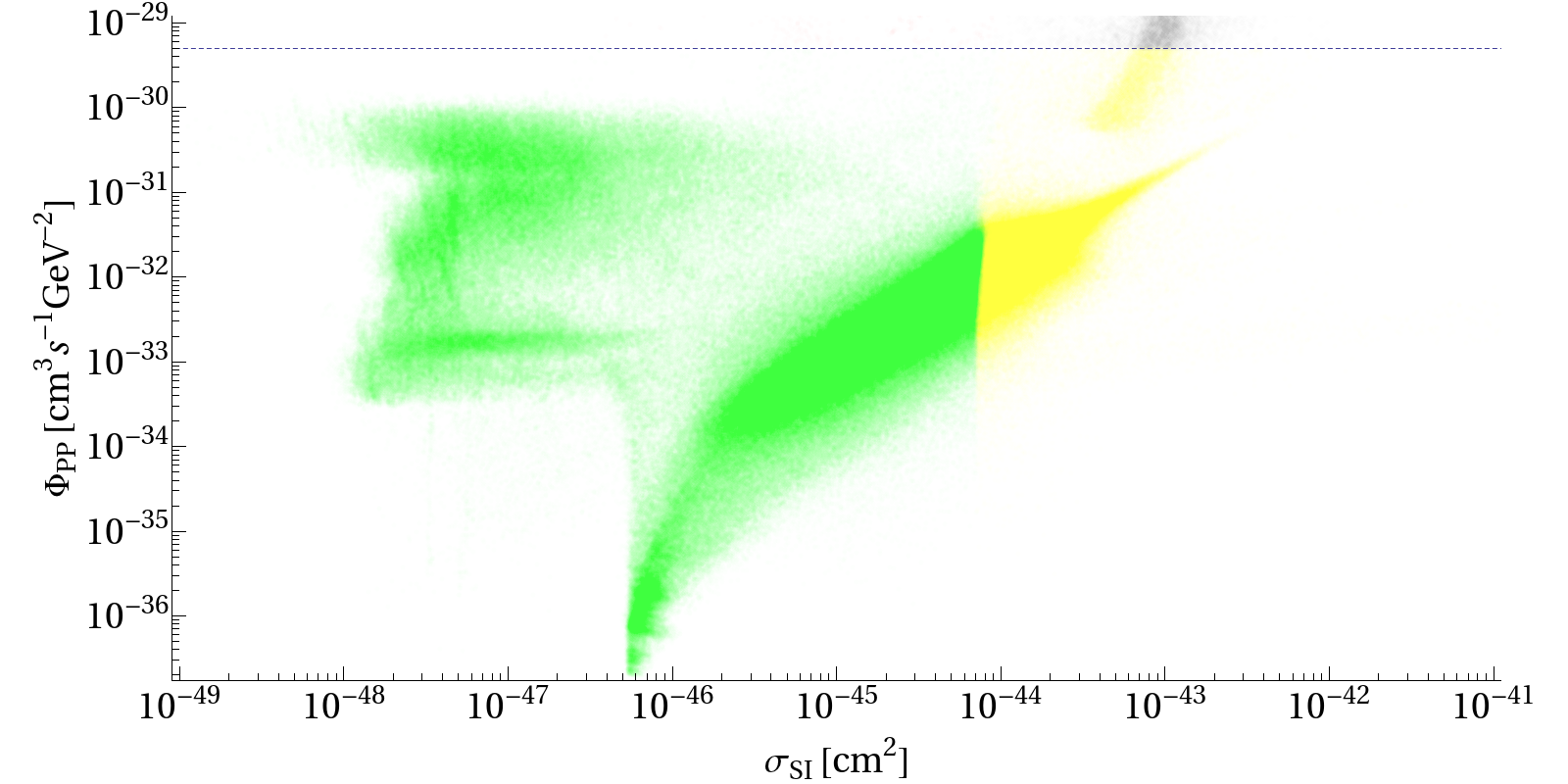}
\caption{$\Phi_{\rm PP}$ vs the spin-independent cross section for $m_{\tilde{\chi}_1^0} < 100$ GeV. The right panel shows the case with the regeneration of the DM relic density  to the correct value, the left panel shows the case without. The limit on $\Phi_{\rm PP}$ shown as a blue dashed line is from the combined analysis of FERMI-LAT observations of dSphs \cite{GeringerSameth:2011iw}. Colour coding is the same as in FIG.~\ref{fig:direct}.}
\label{fig:correlation}
\end{figure*}
In FIG.~\ref{fig:direct} plots of the spin-independent elastic scattering cross section against neutralino mass are shown for Scan A. In these plots the red points are excluded by the constraints from the FERMI-LAT gamma ray limits from dSphs, yellow points are ruled out by XENON100 direct detection searches, grey points are ruled out by both and the green points are those that survive the constraints applied.

In the left panel of FIG.~\ref{fig:direct}\;no regeneration of the DM density is assumed; hence $\eta$ can be small. The result is that for points with a low freeze-out relic abundance, like those in the Z and $h^0$ resonance regions, the elastic scattering cross section can be large, i.e. above the XENON100 limit as evaluated for a DM species with the WMAP observed density, and still predict a sufficiently low event rate in a direct detection experiment to evade the exclusion limits. 

Also visible is a region around and just below $m_{\tilde{\chi}_1^0}\sim100\;$GeV. Comparing with FIG.~\ref{fig:channels_below100GeV}, this region corresponds to the scenarios in which chargino co-annihilations and t-channel exchange diagrams dominate during freeze-out.

If we now assume that the DM density is regenerated after freeze-out to the observed value, all points above the XENON100 limit are now ruled out, as shown in the right panel of FIG.~\ref{fig:direct}. There are a number of points that are still allowed, in particular those that appear in the Z and $h^0$ resonance regions. The reason for this is that if the neutralino DM can annihilate via an on-shell s-channel resonance, the size of the couplings needed to give a large enough annihilation cross section at freeze-out to reduce the DM relic abundance below the WMAP measured value, can be smaller. 

The size of the couplings between the neutralino and both the Z and $h^0$ is determined by the size of the Higgsino component in the neutralino, which in turn determines the size of the spin-independent elastic scattering cross section. This reduction in the couplings will therefore allow some of the points in the resonance regions to avoid the direct detection limit, provided they correspond to points with close to on-shell freeze-out annihilations. Despite this, a significant number of points are ruled out by direct detection.

In FIG.~\ref{fig:RD} we present plots of the distribution of points found by the MCMC in the ($\sigma_{\rm SI}, \Omega_{\rm FO}h^2$) plane. Once again, in the left panel of FIG.~\ref{fig:RD} the DM relic density is kept at the value predicted by freeze-out and in the right the DM density is assumed to have been regenerated to the observed value but is plotted as a function of the relic density generated by freeze-out for each point. 

Different regions of the plots in FIG.~\ref{fig:RD} can be identified and explained in terms of the connection between the annihilation cross section in the early universe and the spin-independent elastic scattering cross section. There are two main regions of points corresponding to different types of process that dominate the DM annihilation cross section at freeze-out, they are, DM annihilation via s-channel $Z$ or $h^0$ and DM co-annihilation with another SUSY particle (usually the chargino). With reference to the left panel of FIG.~\ref{fig:RD}, the points corresponding to s-channel processes are roughly contained within the green diagonal band and the yellow points above. The co-annihilation points are those below the green diagonal band.

Moving from small to large freeze-out abundances (left to right in both panels of FIG.~\ref{fig:RD} but remaining at a constant spin-independent scattering rate, corresponds to moving off-shell for the s-channel annihilation rate at freeze-out. That is, the mass of the neutralino DM is moving away from either $m_Z/2$ or $m_{h^0}/2$. This reduces the overall annihilation rate and therefore increases the freeze-out relic abundance. 

Moving down the plots in FIG.~\ref{fig:RD} we move to smaller spin-independent  elastic scattering cross sections with the size of the Higgsino component in the mostly Bino neutralino decreasing, which results in smaller couplings to $h^0$. The DM s-channel annihilation cross section at freeze-out also decreases with the decreasing couplings and that effect translates into the diagonal slope that can be seen in both plots of FIG.~\ref{fig:RD}. The maximum size of the annihilation cross section at freeze-out, when the s-channel resonance is on shell, decreases with decreasing Higgsino component. Consequently the smallest possible value of the freeze-out relic abundance gets larger as we decrease the spin-independent  elastic scattering cross section leading to the diagonal edge clearly visible in the distribution of points.

The second region corresponding to DM co-annihilations in FIG.~\ref{fig:RD}) has generically lower spin-independent  scattering cross sections but can have a range of relic abundances. The majority of points in this region correspond to situations where the freeze-out process is unrelated to the spin-independent  cross section as is the case for stau co-annihilations and exchange and so no discernible pattern emerges. 

In the left hand panel of FIG.~\ref{fig:RD} the relic abundance is kept at the freeze-out value and the resulting relaxation of the elastic scattering cross section bound is once again apparent due to the reduction of the DM relic density compared to the WMAP observed value. In this scenario the limits from dSphs also play no role due to the suppression in the DM relic density.

In the right hand panel of FIG.~\ref{fig:RD}, with the DM density regenerated to the WMAP observed value, a significant number of points are excluded by direct detection. The effect of the limits from dSphs is quite minimal, only a handful of points (red points in FIG.~\ref{fig:RD}) are ruled out exclusively by this indirect constraint and they are the ones with very low freeze-out relic abundance and hence a large DM annihilation cross section. In particular, these points represent on-shell annihilation through $A^0$.

FIG.~\ref{fig:RD} is particularly interesting as it shows that unless the cross section is very suppressed\footnote{Even with this suppression the number of points in this region is very low.} ($\sigma_{\rm SI} \ll 10^{-44} \rm{cm^2}$), neutralinos with a freeze-out relic density that exceeds one percent of the WMAP upper limit are the only possible type of DM candidates that can be saved via a regeneration mechanism.

In FIG.~\ref{fig:correlation} the same points are shown on plots in the ($\Phi_{\rm PP}, \sigma_{\rm SI}$) plane. These plots give a useful demonstration of the relative importance of the two constraints, with the majority of points being ruled out by direct detection. 

\begin{figure}[t]
	\centering
\includegraphics[width=0.49\textwidth]{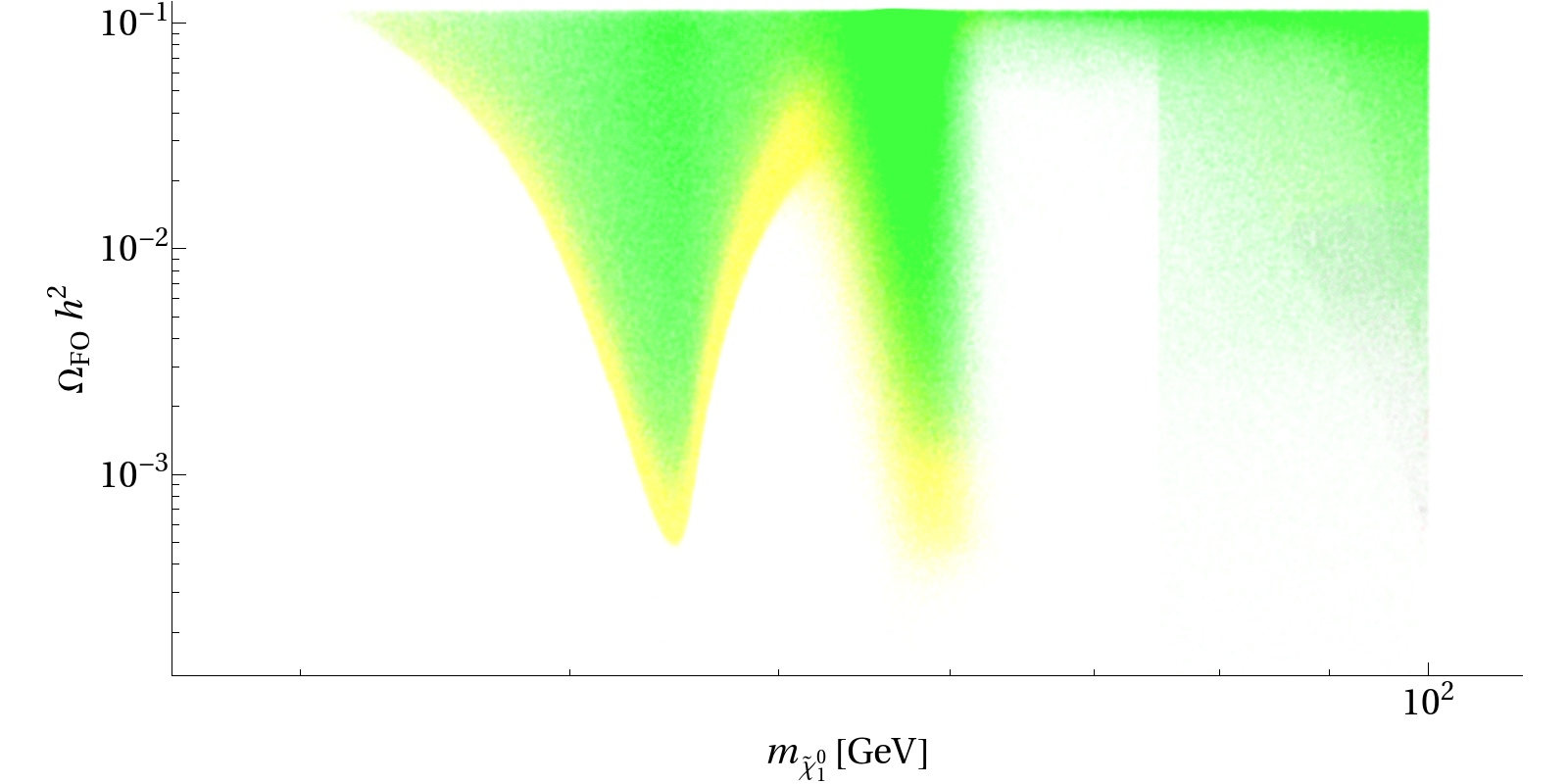}
\caption{A plot of $m_{\tilde{\chi}^0_1}$ versus the freeze-out neutralino relic density where regeneration is assumed. Colour coding is the same as in FIG.~\ref{fig:direct}.}
\label{fig:relicabundance_exclusions_light}
\end{figure}

FIG.~\ref{fig:relicabundance_exclusions_light} shows the final result of applying both direct and indirect detection constraints in the ($m_{\tilde{\chi}^0_1}$, $\Omega_{\rm FO} h^2$) plane assuming the regeneration of the DM relic abundance to the WMAP observed value. It can be seen that indirect detection limits do not constrain the resonant Z and $h^0$ freeze-out annihilation scenarios. Spin-independent direct detection excludes the most under abundant scenarios particularly in the case of resonant annihilation via $h^0$. The interplay between the spin-independent  coupling and resonant effects during freeze-out discussed earlier can again be seen in the thin strip of points excluded around the edges of the Z and $h^0$ resonance regions. It is clear that points further from the resonance regions require  larger couplings in order to reduce the freeze-out relic abundance below the WMAP observed value. This generates a larger spin-independent  cross-section leading to the exclusion of these points by direct detection. 

\subsection{Regeneration in scenarios with $m_{\tilde{\chi}_1^0} >$ 100 GeV}

\begin{figure*}[t]
	\centering
\includegraphics[width=0.49\textwidth]{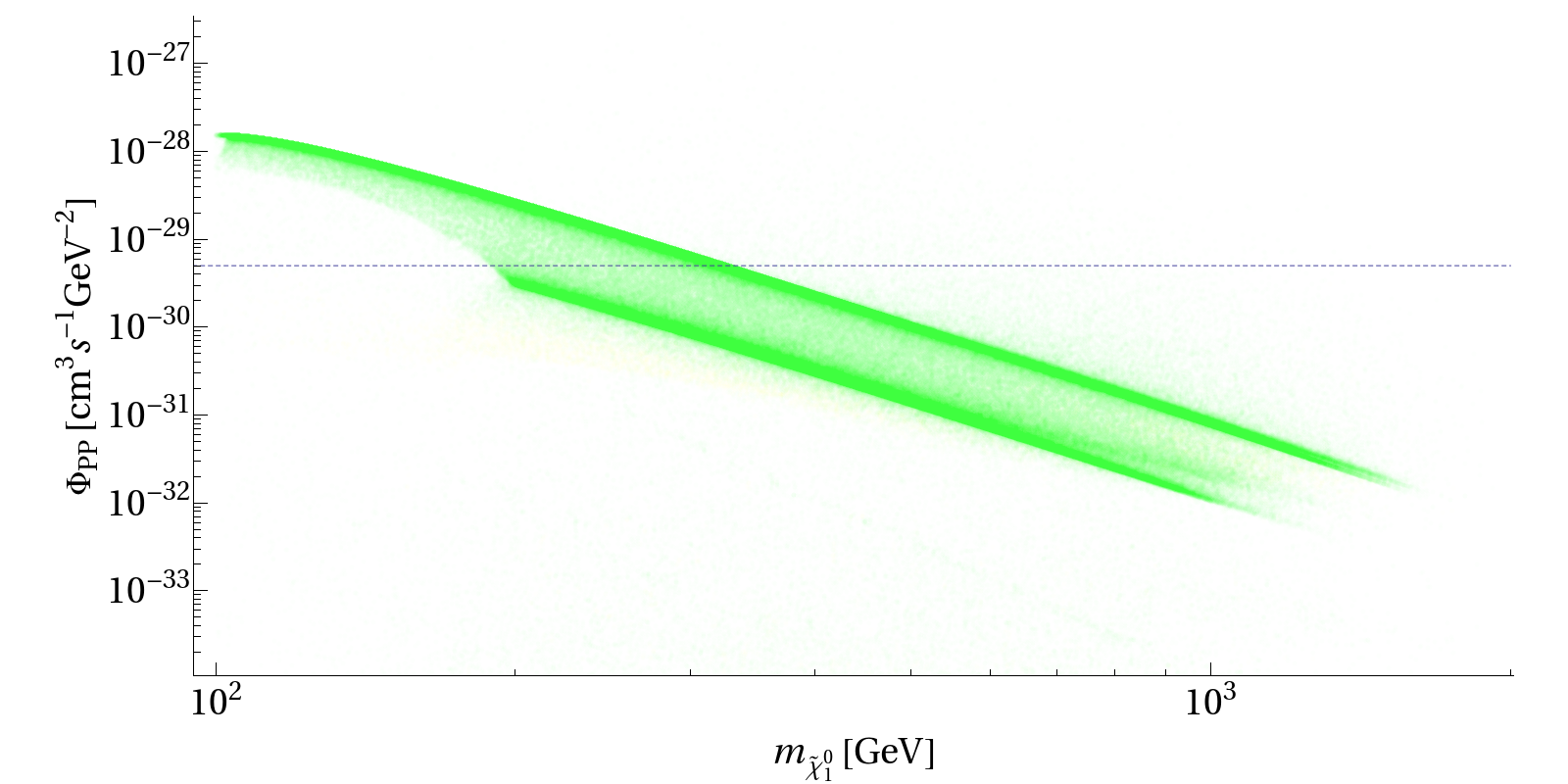}
\includegraphics[width=0.49\textwidth]{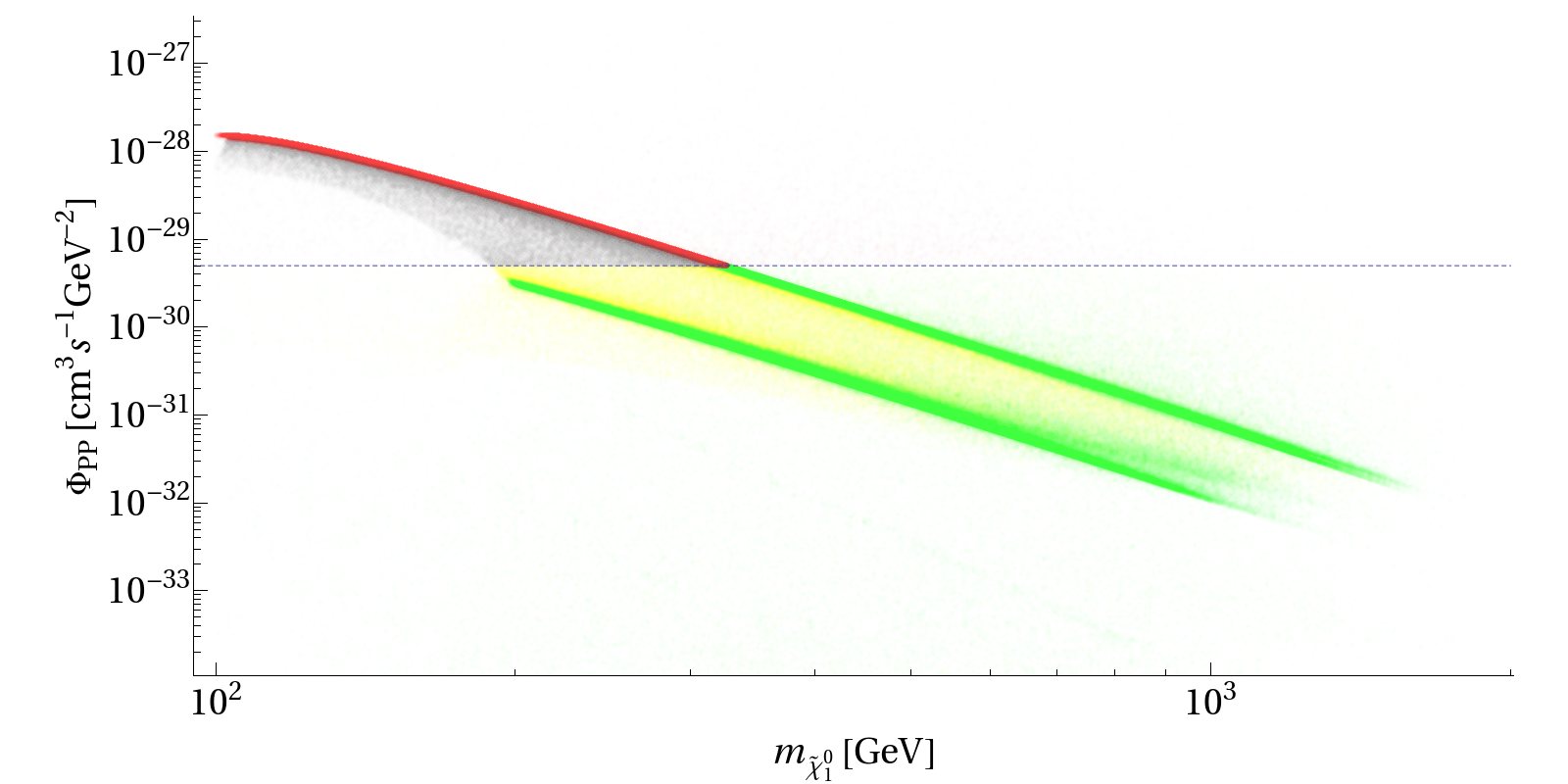}
\caption{$\Phi_{\rm PP}$ vs the neutralino mass for Scan B. The limit shown is from FERMI-LAT observations of dSphs \cite{GeringerSameth:2011iw}. The right panel shows the case with the regeneration of the DM relic density to the correct value, the left panel shows the case without. Colour coding is the same as in FIG.~\ref{fig:direct}.}
\label{fig:dsphheavy}
\end{figure*}

\begin{figure*}[t]
	\centering
\includegraphics[width=0.49\textwidth]{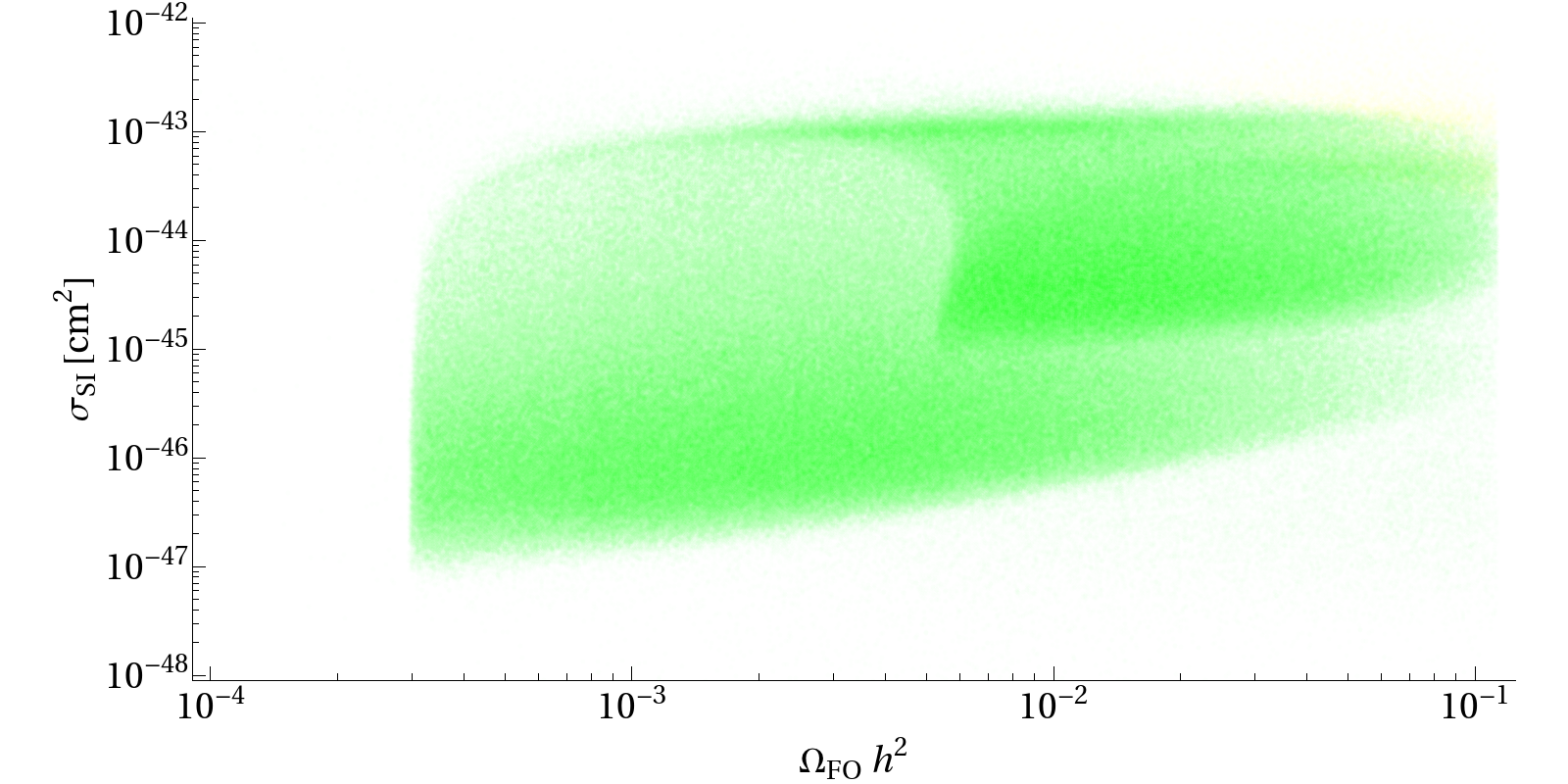}
\includegraphics[width=0.49\textwidth]{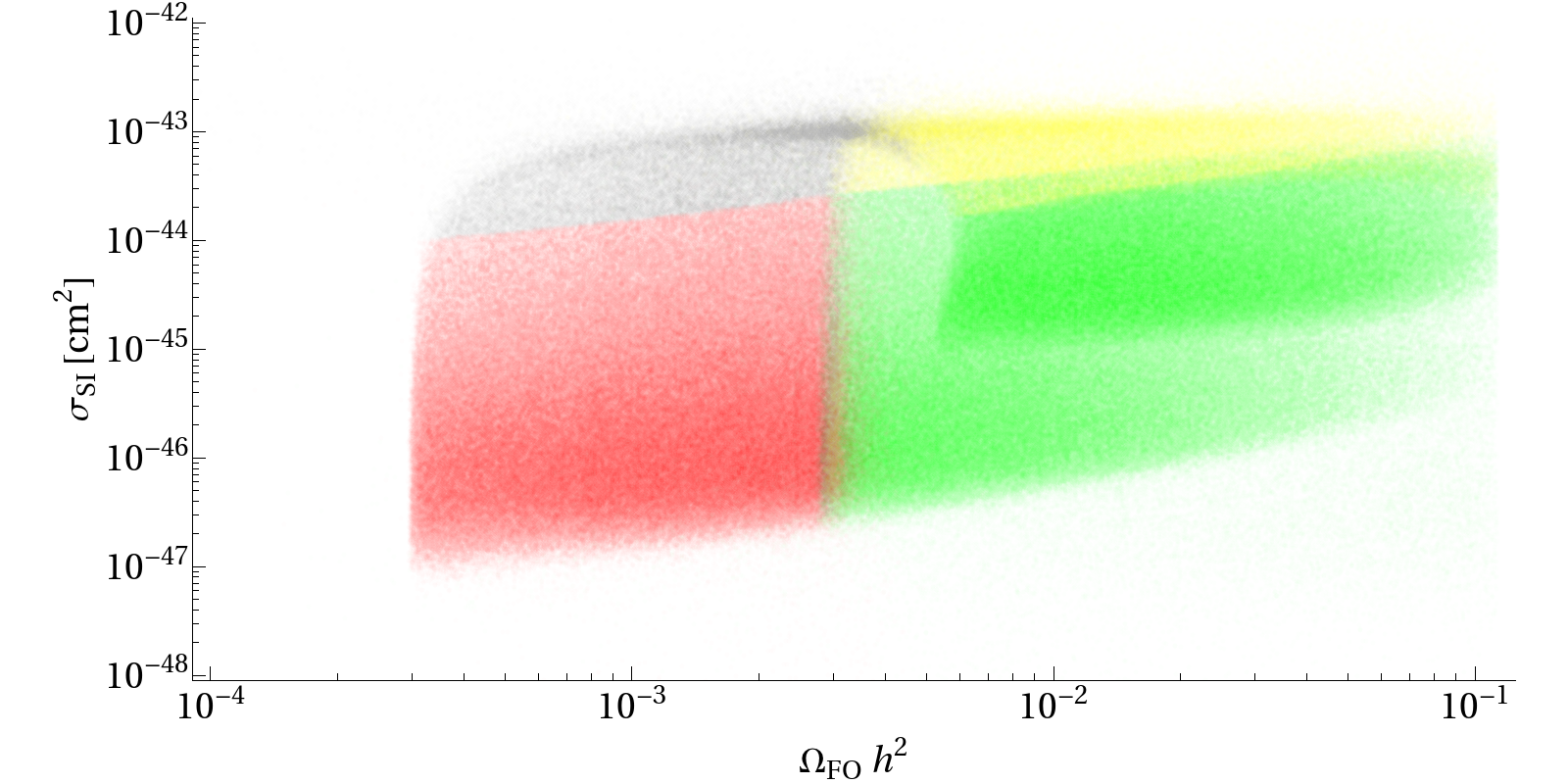}
\caption{Spin-independent cross section versus the neutralino relic density with $m_{\tilde{\chi}^0_1} > 100$ GeV. The right panel shows the case with the regeneration of the DM relic density to the correct value, the left panel shows the case without.  Colour coding is the same as in FIG.~\ref{fig:direct}.}
\label{fig:correlationheavy}
\end{figure*}

Turning now to the heavier candidates of Scan B. FIG.~\ref{fig:dsphheavy} displays the distribution of MCMC points found in the ($\Phi_{\rm PP}, m_{\tilde{\chi}^0_1}$) plane. The colour scheme is identical to the earlier figures with red points ruled out by the dSph limits, yellow points ruled out by the XENON100, grey points ruled out by both and green points are not constrained by either.

The left panel of FIG.~\ref{fig:dsphheavy}, with no regeneration, has no points that are ruled out by the dSph limits. As with neutralinos with masses below $100\;$GeV, the dSph limits plays no significant role in restricting the under-abundant scenarios due to the reduced relic density suppressing the DM annihilation rate into photons.

Regenerating the DM density to the WMAP observed value, the dSph limits now play a significant role in constraining the allowed parameter space as demonstrated in the right panel of FIG.~\ref{fig:dsphheavy}. The points ruled out by the dSph limits correspond to the most under-abundant scenarios, which can be seen clearly in FIG.~\ref{fig:correlationheavy}, which contains plots of points in the ($\sigma_{\rm SI}, \Omega_{\rm FO}h^2$) plane. In the case of regeneration (right panel of FIG.~\ref{fig:correlationheavy}) the impact of the dSph limits is restricted to the most under-abundant scenarios with abundances up to just below $3\%$ of the WMAP observed value being constrained. 

In addition, FIG.~\ref{fig:correlationheavy} shows that direct detection still plays an important role in constraining neutralino DM with masses above $100\;$GeV. In particular, it constrains points with a large range of freeze-out abundances and consequently provides a useful complementary constraint to the dSph limits.

\begin{figure}[t]
	\centering
\includegraphics[width=0.49\textwidth]{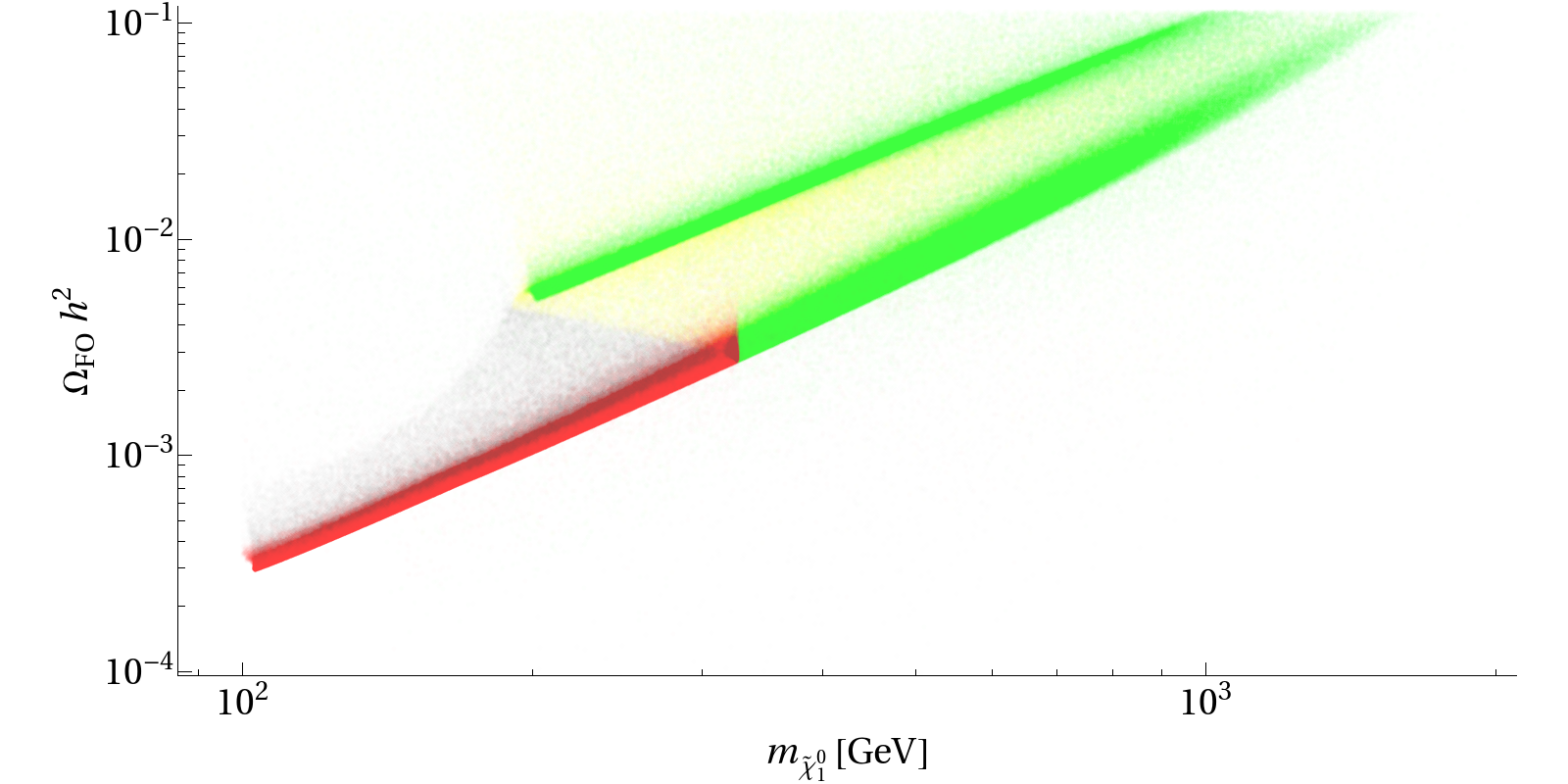}
\caption{Neutralino mass versus the freeze-out neutralino relic density where the regeneration of the DM density is assumed. Colour coding is the same as in FIG.~\ref{fig:direct}.}
\label{fig:relicabundance_exclusions_heavy}
\end{figure}

The result of applying the indirect and direct detection limits in the  ($\Omega_{\rm FO}h^2, m_{\tilde{\chi}^0_1}$) plane after regeneration is shown in FIG.~\ref{fig:relicabundance_exclusions_heavy}. The majority of excluded points come from the lower end of the mass distribution with all points with relic abundances less than around 3\% of the WMAP observed value being ruled out by a combination of the direct detection and dSph limits applied in our analysis. Spin-independent direct detection limits also lead to a reduction of points with larger masses and abundances.

\section{Discusion and Conclusion}\label{conclusions}

Using a familiar framework (neutralinos in the MSSM), we have investigated the configurations for which the expected freeze-out relic density could be much smaller than the upper limit of the WMAP observed  value. We have found many configurations where $\Omega_{\rm FO} h^2$ could be down to $10^{-5} \Omega_{\rm WMAP} h^2$. In particular, for low neutralino masses, resonant annihilation through Higgs or $Z$ boson appear to be very efficient. 

However, we have demonstrated that if a mechanism is capable of regenerating the candidate DM number density to the present observed value, the combination of FERMI-LAT gamma ray observations in dSph and DM direct detection limits from XENON100 make such scenarios difficult to realise, thereby suggesting that candidates with very small freeze-out relic density (less than a percent of the WMAP upper limit) cannot be the sole explanation to the DM  problem even if one assumes that after freeze-out the DM density is regenerated.

One of the central points of this study is the link between the DM annihilation process at freeze-out and the predicted direct detection rate. Essentially, as is already known in the MSSM scenario a light Higgs or $Z$ exchange is needed in order for light neutralinos to become extremely under-abundant.  Small neutralino 
freeze-out abundances correspond to scenarios which are close to the Higgs resonance and thus typically predict large neutralino-nucleon elastic scattering cross sections since one is close to the minimal Higgs mass value that is possible in the MSSM (given collider constraints) and the neutralino-Higgs couplings are constrained by the requirement of a large annihilation cross section which is itself bounded by the Higgs decay width\footnote{For scenarios which lie very close to the resonance the couplings cannot be arbitrarily small since the decay width will dominate over a very small mass degeneracy between the Higgs and the neutralino.}. 

In scenarios without regeneration, such a large neutralino-nucleon elastic scattering cross section (or large annihilation cross section) is not necessarily excluded. Indeed, the very small freeze-out abundance actually induces a '$\eta$' suppression factor and reduces the elastic scattering rate in nuclear recoil direct detection experiments as well as the indirect detection rate. However in scenarios where one allows regeneration of the relic density to happen, there is no '$\eta$' suppression factor and these scenarios can be ruled out by direct and indirect detection experiments. 

The heavy neutralino scenarios which are under-abundant (less than 3$\%$ of the observed relic density) also benefit from the '$\eta$' factor suppression if there is no regeneration mechanism involved, so they cannot be ruled out. However, when we assume regeneration, we find that the indirect detection constraint set by the FERMI-LAT experiment actually rule out these candidates and complement the constraint set by XENON100.

Our conclusion is based on the combination of astrophysics, astroparticle and particle physics data. Any more constraints in, at least, one of these fields will enable stronger constraints to be set, thus restricting the types of mechanisms that could give the DM the relic density it has today.

In this analysis we have not applied the latest limits emerging form the LHC on the sparticle spectrum. Applying these limits will reduce the number of allowed points further making the regeneration scenario even harder to realise. We leave this study to future work.

\acknowledgments
SMW and AJW thank the Higher Education Funding Council for England and the Science and Technology Facilities Council for financial support under the SEPNet Initiative and the IPPP for financial support.

\bibliography{SusyFI_00}

\end{document}